\documentclass[useAMS,natbib]{mn2e}
\usepackage{graphicx}
\usepackage{rotating,times,pictex,graphicx,latexsym}
\usepackage{color}
\usepackage{longtable,amsmath}
\usepackage{lscape}

\title{Old Star Clusters in the FSR catalogue}  

\author[Froebrich et al.]{D. Froebrich$^{1}$\thanks{E-mail: df@star.kent.ac.uk},
S. Schmeja$^{2}$, D. Samuel$^{3}$, P.W. Lucas$^{3}$\\ $^1$ Centre for
Astrophysics and Planetary Science, University of Kent, Canterbury, CT2 7NH, UK
\\ $^2$ Zentrum f\"ur Astronomie der Universit\"at Heidelberg, Institut f\"ur
Theoretische Astrophysik, Albert-Ueberle-Str. 2, 69120 Heidelberg, Germany \\
$^3$ Centre for Astrophysics Research, University of Hertfordshire, College
Lane, Hatfield, AL10 9AB, UK} 

\begin{document}

\date{Received sooner; accepted later}
\pagerange{\pageref{firstpage}--\pageref{lastpage}} \pubyear{2007}
\maketitle

\label{firstpage}

\begin{abstract}

We investigate the old star clusters in the sample of cluster candidates from
Froebrich, Scholz \& Raftery 2007 -- the FSR list. Based on photometry from the
2\,Micron All Sky Survey we generated decontaminated colour-magnitude and
colour-colour diagrams to select a sample of 269 old stellar clusters. This
sample contains 63 known globular clusters, 174 known open clusters and 32 so
far unclassified objects. Isochrone fitting has been used to homogeneously
calculate the age, distance and reddening to all clusters. The mean age of the
open clusters in our sample is 1\,Gyr. The positions of these clusters in the
Galactic Plane show that 80\,\% of open clusters older than 1\,Gyr have a
Galactocentric distance of more than 7\,kpc. The scale height for the old open
clusters above the Plane is 375\,pc, more than three times as large as the
115\,pc which we obtain for the younger open clusters in our sample. We find
that the mean optical extinction towards the open clusters in the disk of the
Galaxy is 0.70\,mag/kpc. The FSR sample has a strong selection bias towards
objects with an apparent core radius of 30" to 50" and there is an unexplained
paucity of old open clusters in the Galactic Longitude range of 120$^\circ < l
<$\,180$^\circ$.

\end{abstract}


\section{Introduction}

As birthplaces for the majority of stars (e.g. Lada \& Lada
\cite{2003ARA&A..41...57L}) stellar clusters can be considered the building
blocks of galaxies. The vast majority of them only reaches ages of a few Myrs
after which their member stars dissolve into the general field star population.
The disruption timescales are dependent e.g. on the local tidal gravitational
field (interaction with nearby giant molecular clouds), the star formation
efficiency in the cluster, the mass of the cluster and the efficiency of the
feedback from the young stars in the cluster (jets, winds, supernova
explosions). There is evidence that the disruption timescales are increasing
with distance from the Galactic Center (e.g. Lamers \& Gieles
\cite{2006A&A...455L..17L}, Goodwin \& Bastian \cite{2006MNRAS.373..752G},
Piskunov et al. \cite{2007A&A...468..151P}). A number of clusters, however,
survive this initial infant mortality phase and become open clusters, which then
can reach ages of up to several Gyrs.

These old stellar systems, including both, open and globular clusters, provide
us with laboratory like conditions. All stars within such a cluster can be
considered as being situated at the same distance, having the same age and
metallicity. Due to their age, they are usually not associated with giant
molecular clouds, thus there is a constant reddening towards all cluster
members. Hence, one can fit theoretical isochrones to the cluster
colour-magnitude diagram to determine the age, distance and reddening
simultaneously, provided the metallicity is known. 

As current catalogues of old open clusters are rather incomplete (e.g. Bonatto
\& Bica \cite{2007A&A...473..445B}), our aim is to establish a large, well
defined sample of such old stellar systems and to determine its properties in a
homogeneous way. This will then be used to investigate the distribution of these
old clusters in the Galaxy which will improve our understanding not just of the
old stellar systems, but also on issues such as the interstellar extinction law,
disruption timescales of clusters, and ultimately the chemical evolution and
enrichment history of the Galactic Disk.

To obtain a large sample of old clusters and analyse its properties
homogeneously, we utilise the 2 Micron All Sky Survey (2MASS, Skrutskie et al.
\cite{2006AJ....131.1163S}) point source catalogue and the star cluster
candidate list provided by Froebrich et al. \cite{2007MNRAS.374..399F}. We
identify the old systems amongst their catalogue by investigation of
decontaminated colour-magnitude and colour-colour diagrams and determine their
parameters by fitting theoretical isochrones from Girardi et al.
\cite{2002A&A...391..195G} to the 2MASS photometry.

Our paper is structured as follows. In Sect.\,\ref{dataanalysis} we describe the
selection of our cluster sample and the determination of its properties. This
includes the automatic decontamination of foreground stars in the cluster
fields, the selection and identification of the old stellar systems and the
determination of their ages, distances and reddening via isochrone fits. In
Sect.\,\ref{results} we present our main results and discussion. We characterise
the cluster sample, discuss the distribution of clusters in the Galactic Plane
and identify selection effects. Finally we present our conclusions in
Sect.\,\ref{conclusions}.

\section{Data Analysis}\label{dataanalysis}

\subsection{The FSR sample}

The sample of clusters analysed in this work is based on the FSR catalogue by
Froebrich et al. \cite{2007MNRAS.374..399F}. They determined a star density map
based on 2MASS data (Skrutskie et al. \cite{2006AJ....131.1163S}) along the
entire Galactic Plane with $|b| < 20^\circ$. Star cluster candidates were
selected as local star density enhancements and a total of 1788 objects were
found. These candidates were cross referenced with the SIMBAD database. This
uncovered that the FSR list contained 86 known globular clusters, 681 known open
clusters and 1021, so far unknown cluster candidates. An estimate of the
contamination suggested that about half of these new candidates are real star
clusters. A number of these have been confirmed as real clusters since then. See
Froebrich et al. \cite{2008MNRAS.390.1598F} for a recent summary of FSR cluster
candidates investigated so far.

\subsection{Accurate cluster positions and radii}

Our first aim was to determine more accurate cluster coordinates and the radius
for each FSR cluster candidate. We hence extracted the 2MASS photometry for all
stars in a 0.5$^\circ \times$\,0.5$^\circ$ sized field around each cluster. Only
stars with reliable photometry (quality flag A to C in each of the JHK bands;
Skrutskie et al. \cite{2006AJ....131.1163S}) were used. We then modelled the
cluster candidates by two-dimensional angular Gaussian distributions applying an
expectation-maximization algorithm (Dempster, Laird \& Rubin \cite{dempster77})
and evaluating the best fit using the Bayesian information criterion (BIC;
Schwarz \cite{schwarz78}) by means of a code developed for the cluster search in
UKIDSS GPS data (Samuel \& Lucas, in preparation). This procedure provides us
with the cluster centre and the size of the best-fit Gaussian and the BIC value
-- essentially a description of how probable it is that a given cluster
candidate is a real star cluster. Objects with a BIC value less than zero are
generally considered real, and a smaller BIC value indicates a higher
probability to be a real cluster. The obtained central coordinates, and BIC
values for each of the investigated clusters are listed in
Table\,\ref{properties}.

With the more accurate central positions for each cluster candidate we
calculated radially averaged star density profiles $\rho(r)$. Those profiles
were fit automatically to the function:

\begin{equation}\label{eqradius}
\rho(r) = \rho_{bgr} + \frac{\rho_{cen}}{1 + \left( \frac{r}{r_{core}}
\right)^2},
\end{equation}
where $\rho_{cen}$ and $\rho_{bgr}$ are the central cluster and background star
densities and $r_{core}$ the radius of the cluster. Using the distances to the
clusters, we later convert these radii into real sizes in parsec.

\subsection{Membership probabilities}

To determine the cluster properties via isochrone fitting we need to identify
the most likely cluster member stars. This is in particular important in the
high star density fields near the Galactic Plane, where field star contamination
is important. We used the position and radius of each cluster to define a
cluster region and a control field in the 0.5$^\circ \times$\,0.5$^\circ$ area
around the cluster coordinates. Stars which were closer than three times the
cluster radius to the centre are considered part of the cluster area, all stars
further away than five times the cluster radius are part of the control field. 

We then applied a variation of the colour-colour-magnitude (CCM) decontamination
procedure from Bonatto \& Bica \cite{2007MNRAS.377.1301B} to the stars in the
cluster area. For each star $i$ with the apparent 2MASS magnitudes $J^i$,
$H^i$, $K^i$ and colours $J^i - H^i = JH^i$, $J^i - K^i = JK^i$ we calculate the
CCM distance $r_{ccm}$ to every other star $j \ne i$ in the following way:

\begin{equation}
r_{ccm} = \sqrt{ \frac{1}{2} \left[ J^i - J^j \right]^2 + \left[ JK^i - JK^j
\right]^2 + \left[ JH^i - JH^j \right]^2 } 
\end{equation}

The factor of 0.5 in front of the differences in the J-band magnitudes accounts
for the generally larger spread of the magnitudes compared to the colours. We
determine $r_{ccm}^{10}$ as the 10$^{th}$ smallest value over all stars $j \ne
i$ and count the number $N_{ccm}$ of stars in the control area that are within
the CCM distance $r_{ccm}^{10}$ around the values $J^i$, $JK^i$ and $JH^i$. The
probability $P_{ccm}$ of star $i$ to be a member of the cluster is then given
by:

\begin{equation}
P_{ccm} = 1.0 - \frac{N_{ccm}}{10} \frac{A_{cl}}{A_{con}} 
\end{equation}

where $A_{cl}$ and $A_{con}$ are the areas of the cluster and control
field, respectively. If $P_{ccm}$ for a particular star is negative, then its
membership probability is zero. This approach, instead of a fixed CCM cell size
as in Bonatto \& Bica \cite{2007MNRAS.377.1301B} gives better results for the
probabilities in regions of the CCM space with only a few stars, i.e. at bright
magnitudes.

Alternatively one could determine the probability $P_{pos}$ for each star to be
a cluster member based on its distance from the cluster centre by assuming that
the projected cluster star density profile has a given distribution $\rho_{\rm
cl}(r)$ (e.g. a Gaussian or similar to Eq.\,\ref{eqradius}), overlayed on a
constant background star density $\rho_{\rm bgr}$. Stars outside five times the
cluster radius could be used to determine the background star density $\rho_{\rm
bgr}$. Based on the distance $r$ of each individual star to the cluster centre,
one could estimate its probability $P_{pos}(r)$ to be a cluster member based of
the star density at this position via: 

\begin{equation}
P_{pos}(r) = \rho_{\rm cl}(r) / \rho_{\rm bgr}.
\end{equation}

Both probabilities $P_{ccm}$ and $P_{pos}$ could be combined to a total
membership probability $P = \sqrt{ P_{ccm} \cdot P_{pos}}$. However, we find
that using the position does not give reliable results in many cases. In
particular in dense (globular) clusters, where no stars are detected in 2MASS in
the cluster centres, the probabilities are not reliable. Furthermore, for
clusters in regions of high background star density the membership probability
for most stars will drop below 20\,\%, despite the fact that their colours are
clearly different from the field. Hence, for the purpose of this paper we solely
use the membership probabilities of stars determined from the CCM
considerations.

\subsection{Selection of old star clusters}

Utilising the individual membership probabilities for all stars in each cluster 
we plotted J-K vs. K colour-magnitude (CMD) and H-K vs J-H colour-colour (CCD)
diagrams for each FSR cluster candidate. In Fig.\,\ref{examplecmd0412} we show
the diagrams, including the best fitting isochrone (see below) of the so far
uninvestigated cluster FSR\,0412 (Pfleiderer\,3) as an example. One can nicely
see that the cluster red giant stars are the most likely members ($P_{ccm} >
80$\,\%). Stars possessing colours in agreement with foreground dwarf stars are
much less likely to be cluster members. In Appendix\,\ref{CMD} we show the CMDs
and CCDs for all clusters investigated in this paper. 

\begin{figure}
\includegraphics[width=6.1cm,angle=-90]{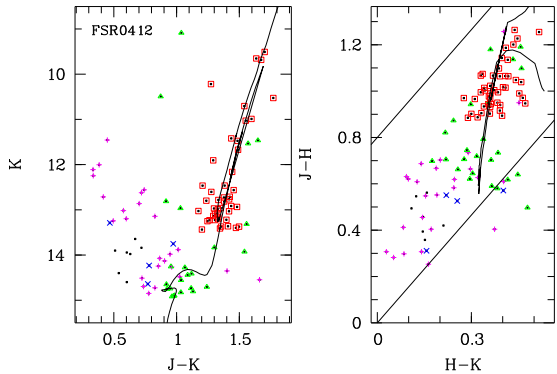}
\caption{\label{examplecmd0412} Example of our colour-magnitude (left) and
colour-colour diagrams for the cluster FSR\,0412 (Pfleiderer\,3) which has so
far not been investigated in detail. Red squares are stars with
$P$\,$>$\,80\,\%, green triangles are stars with 60\,\%\,$<$\,$P$\,$<$\,80\,\%,
pink $+$-signs are stars with 40\,\%\,$<$\,$P$\,$<$\,60\,\%, blue crosses are
stars with 20\,\%\,$<$\,$P$\,$<$\,40\,\% and black dots are stars with
$P$\,$<$\,20\,\%. Overplotted in black is the best fitting isochrone
(d\,=\,6.1\,kpc, log(age/yr)\,=\,9.1, A$_K$\,=0.46\,mag and solar
metallicity). The two solid lines in the right panel enclose the reddening band
for stellar atmospheres.}
\end{figure}

We then inspected the 1788 CMDs and CCDs generated for the entire FSR catalogue,
to decide if the high probability members are consistent with a sequence
representing an old stellar cluster. In other words, we manually selected all
FSR objects that either showed a Red Giant Branch (RGB) or the top of the Main
Sequence (MS) and a number of giant stars. Note that this selection has been
performed 'blind', without the knowledge of which object is which cluster (known
or unknown) in order to ensure an unbiased selection. In total 269 of the 1788
objects were selected as candidates for old clusters and analysed in more
detail for this paper. 

\subsection{Identification of known old star clusters}

We cross-identified the list of 269 clusters with the
SIMBAD\footnote{http://simbad.u-strasbg.fr/simbad/sim-fid} database. In total 63
known globular clusters are in the list, 174 known open clusters (including some
already confirmed FSR objects) and 32 so far unclassified FSR cluster
candidates. Some obviously old clusters, in particular some of the known
globular clusters (e.g. FSR\,0005 or NGC\,6569, vdB-Hagen\,260), are missing in
our sample of old FSR clusters. This is mainly caused by the fact that they do
not contain a large enough number of high probability cluster members,
representing an old stellar sequence.

We obtained the distances, metallicities and reddening for the known globular
clusters from the list of Harris \cite{1996AJ....112.1487H}. The parameters for
FSR\,0040 (2MASS\,GC\,1) are obtained from Ivanov et al.
\cite{2000A&A...362L...1I} and the values for FSR\,1735 are taken from Froebrich
et al. \cite{2008MNRAS.390.1598F}. The cluster FSR\,1762 (Pismis\,26,
vdB-Hagen\,71, Tonantzintla\,2) is listed as globular or cluster of stars in
SIMBAD and we used its parameters from the list of Harris
\cite{1996AJ....112.1487H}. The clusters FSR\,0190, 0584 and 1716 are also
listed as globular or open cluster. In those cases we used the literature data
from Froebrich et al. \cite{2008MNRAS.383L..45F}, Bica et al.
\cite{2007A&A...472..483B} and Froebrich et al. \cite{2008MNRAS.390.1598F} and
Bonatto \& Bica \cite{2008A&A...491..767B}, respectively.

The open cluster parameters were obtained (as first choice) from the
WEBDA\footnote{http://www.univie.ac.at/webda/} database for galactic open
clusters. If no data was available for an open cluster we searched the
literature. The main Table\,\ref{properties} with the cluster parameters
indicates the papers used in those cases. In total we obtained data for 147 of
the known open clusters. For 27 open clusters no data was available and their
properties have hence been determined here, together with the parameters for the
32 so far unclassified FSR cluster candidates.

\subsection{Cluster parameter determination}

From our analysis so far we only determined the cluster position and radius, as
well as the BIC value. In order to determine the cluster parameters such as
distance, reddening and age, we need to fit an appropriate isochrone to the CMD
and CCD for each cluster. We used the isochrone models from Girardi et al.
\cite{2002A&A...391..195G} for 2MASS data to perform this task. The Figures
containing the CMDs and CCDs for all selected old FSR clusters in
Appendix\,\ref{CMD} show in general two isochrones: One with the literature
values for the cluster and our best fitting isochrone. The literature isochrone
is shown as dashed blue line, the best fitting isochrone from this paper is
shown as a solid black line. The parameters used for our best isochrone fit for
all clusters are listed in Table\,\ref{properties}. The uncertainties of the
determined parameters are discussed in Sect.\,\ref{comp}.

The reddening to each cluster used in our best fitting isochrone is given as the
K-band extinction in Table\,\ref{properties}. To overplot the isochrones on the
CMDs and CCDs we need to convert the K-band into the J and H-band extinction
using $A_J = C_{JK} * A_K$ and $A_H = C_{HK} * A_K$. We use a conversion factor
$C_{JK} =$\,2.618 following Mathis \cite{1990ARA&A..28...37M}. In order to fit
the isochrone data in the CCD as well, in general the conversion factor $C_{HK}
=$\,1.529 from Mathis \cite{1990ARA&A..28...37M} seems too low. For the majority
of clusters we hence use $C_{HK} =$\,1.67. However, in some cases those values
do not provide a satisfying fit, and we hence adjusted the value for $C_{HK}$
for each cluster separately. The used values for each cluster are listed in
Table\,\ref{properties}.

\section{Results and Discussion}\label{results}

\subsection{General}

We have identified 269 old stellar clusters in the FSR catalogue of possible
cluster candidates. For the 63 known globular clusters and 147 known open
clusters we extracted parameters (distance, reddening, age) from the literature.
For the remaining 27 known open clusters and 32 so far unclassified FSR cluster
candidates we determine parameters here using isochrone fitting. Additionally,
we determine the parameters of all clusters homogeneously by the same set of
data (2MASS JHK photometry), the same data analysis method and the same set of
isochrones (Girardi et al. \cite{2002A&A...391..195G}). This will allow us, in
particular for the open clusters, to analyse and compare the distribution of the
parameters of our sample of old stellar clusters along the entire Galactic
Plane. We will in the following only discuss the open cluster parameters, if not
stated otherwise.

In Appendix\,\ref{notes} we provide some notes for all newly identified old open
clusters and for the known ones when their parameters differ significantly from
the literature values. Here we will briefly mention some of the notable
discoveries and their properties. FSR\,0039 is a 1\,Gyr old, highly reddened
cluster. With just 4.6\,kpc from the Galactic Centre it is one of the rare old
inner Galaxy clusters. FSR\,0313 (Kronberger\,81) shows a large number of giants
but no main sequence. This indicates that it might be an old, massive cluster
about 10kpc from the Galactic Centre. Both, FSR\,0412 (Pfleiderer\,3) and
FSR\,0460 are very distant (6.1\,kpc) and old (1.2\,Gyr) clusters. Very nice
examples of newly discovered clusters (or objects with parameters determined for
the first time) are FSR\,0134, 0177 (Kronberger\,52), 0275, 0342, 0972
(NGC\,2429), 1404 (vdB-Hagen\,55), 1463, 1565 (Trumpler\,19), 1670 (Loden\,1101)
which show a number of red giants and main sequence stars, while for FSR\,0170,
0329 (Berkeley\,92), 1521, 1559 (Teutsch\,106) only red giants are detected.

\subsection{Age distribution}

\begin{figure}
\includegraphics[width=6.1cm,angle=-90]{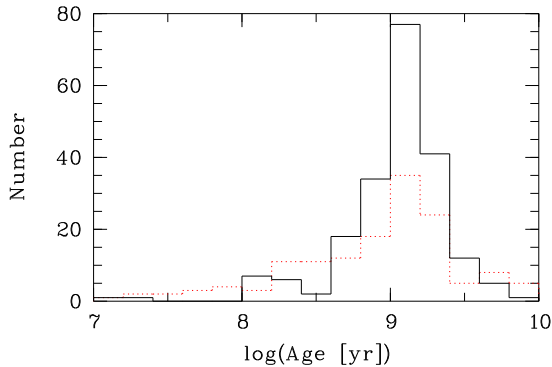}

\caption{\label{distrage} Distribution of the ages of the open clusters in our
sample. The red dotted histogram shows the distribution of ages (from the
literature) for the known open clusters. The black solid histogram shows the
distribution for the ages (determined in this paper) of all open clusters in our
sample. Typical age uncertainties are within the bin width of the histogram (see
also Sect.\,\ref{comp}).}

\end{figure}

Since we aim to investigate the old clusters in the FSR catalogue, we have to
analyse the age distribution of our sample. This is shown in
Fig.\,\ref{distrage}. There the red dotted histogram shows the age distribution
as obtained for the known open clusters using the literature values. The black
solid histogram shows the distribution of the ages determined in the paper for
all open clusters. In both cases there is a clear peak at about 1\,Gyr (which is
also the average of the distribution), and more than 80\,\% of the clusters are
older than 500\,Myrs. This shows that our selection of clusters with a clear RGB
or a main sequence and red giants, was successful in picking out old stellar
systems. However, it still selects a few younger clusters. Most likely these are
more massive, hence showing a larger, and thus observable number of red (super)
giants earlier in their evolution. Some of the clusters with lower ages (based
on the literature) have, according to our isochrone fits, an older age. This is
caused by the fact that we try to include potential giant stars in the fit of
the cluster isochrone, generally leading to a slightly larger age. 

\subsection{Comparison with literature data}\label{comp}

For a large fraction of clusters we can compare our determined parameters with
the values obtained from the literature. This will allow us to estimate the
uncertainties of the isochrone fitting for the clusters without known
parameters. 

At first we check the position accuracy of the cluster candidates. We determine
the difference of our coordinates and the literature coordinates for the known
clusters. For the generally highly concentrated globular clusters the average
difference is 0.5', while for the open clusters we find an average
positional difference of 2'. This rather large value seems to be caused by
erroneous coordinates of some not well investigated clusters in SIMBAD. See
Table\,\ref{properties} to check the differences for each individual cluster. 

Except in some cases the distance, age and reddening estimates from the
literature and our isochrone fitting are in agreement. In the
Appendix\,\ref{notes} we will discuss in detail the clusters with large
differences in the parameters. On average the cluster distances show a scatter
of about 30\,\% between the literature values and our estimates. For the
log(age) values an agreement of about 10\,\% is found. The reddening values also
agree to within 30\,\%. The 2MASS photometry does not allow the determination of
the metallicity. Hence, we generally used solar values, except if a different
value was available from the literature. In some cases it was, however, only
possible to obtain a fit to the CMDs and CCDs with non-solar values. See
Table\,\ref{properties} for the metallicities used for our best fitting
isochrone. Please note that if the cluster has a lower metallicity than used
here, the estimated reddening would be higher and the distances lower. Similarly
the cluster age would be influenced systematically. However, if the metallicity
is changed by less than a few tenth of a dex, then the parameters will stay
within the above mentioned uncertainties. It is much more important to identify
the cluster red giants and main sequence turn off with high accuracy.

\subsection{Distribution in the Galactic Plane}

 %
 %

\begin{figure*}
\includegraphics[width=6.1cm,angle=-90]{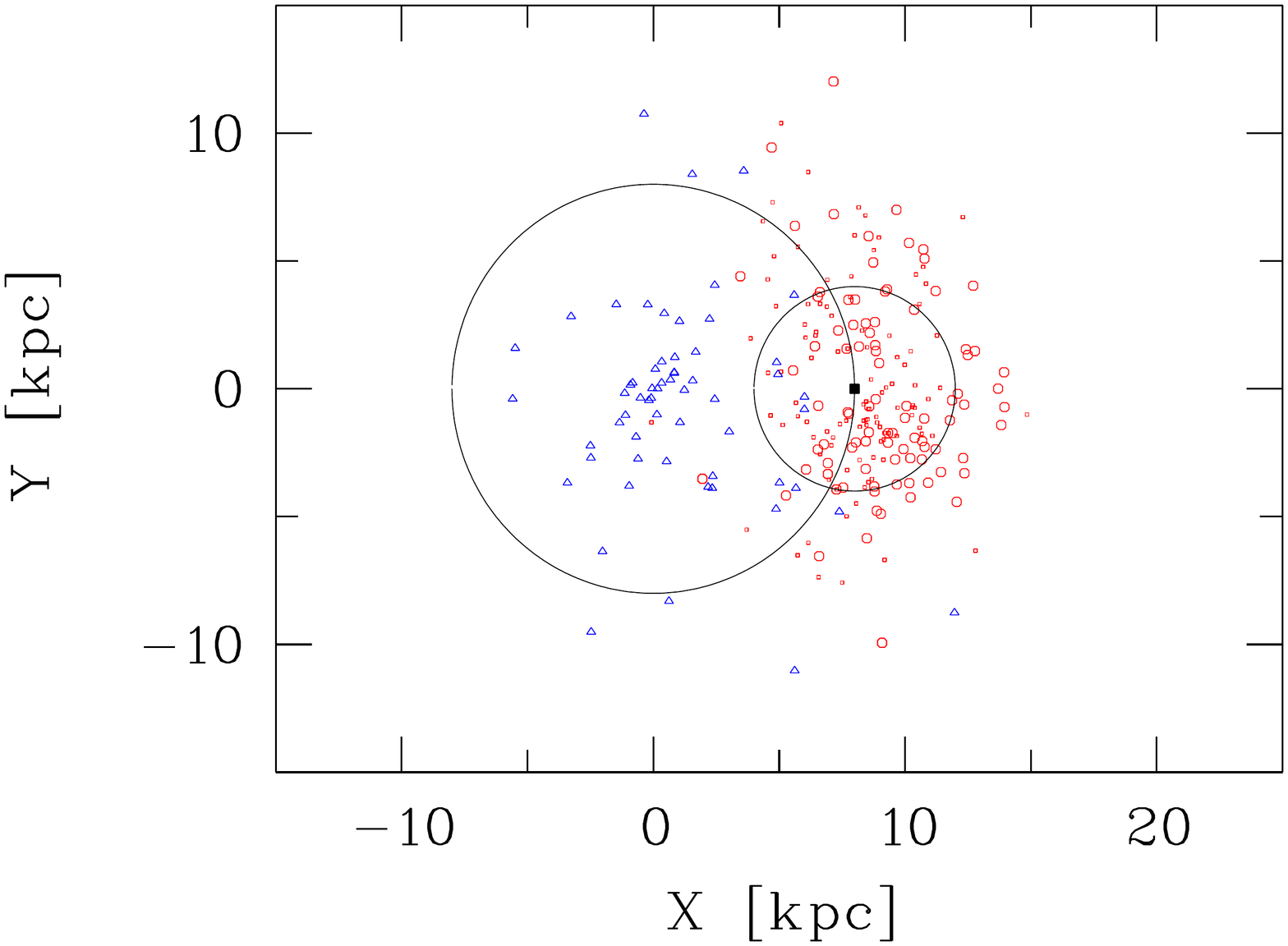}\hfill
\includegraphics[width=6.1cm,angle=-90]{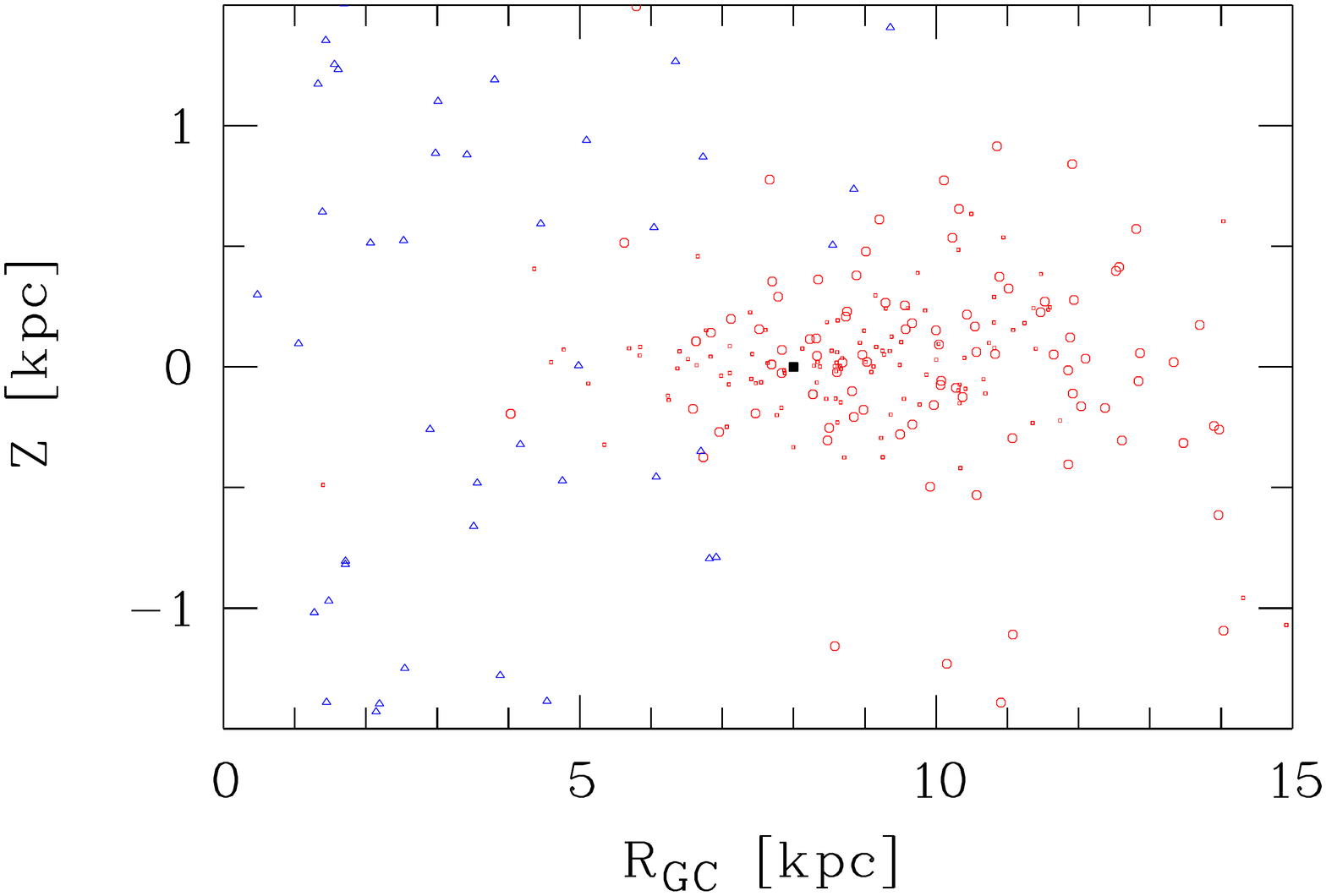}

\caption{\label{distrxy} {\bf Left:} Distribution of the clusters in the
Galactic Plane based on the distances determined in this paper. Known globular
clusters are shown as blue triangles, old open clusters (age above one Gyr) as
red circles and younger open clusters as red dots. The distance of the Sun to
the Galactic Centre is assumed to be 8\,kpc and the Suns position is indicated
by the black square. The two circles indicate a distance of 8\,kpc (large
circle) from the Galactic Centre and 4\,kpc (small circle) from the Sun. Note
that the three globular clusters FSR\,0021 (M\,54), FSR\,0164 (NGC\,7006) and
FSR\,1745 (Terzan\,3) are outside the plotted area. {\bf Right:} Distribution of
the clusters perpendicular to the Galactic Plane of the Galaxy. Shown is the
height Z above the Plane against the Galactocentric distance. The same symbols
as in the left panel are used and for clarity only the region containing the
open clusters is shown.} 

\end{figure*}

\begin{figure}
\includegraphics[width=6.1cm,angle=-90]{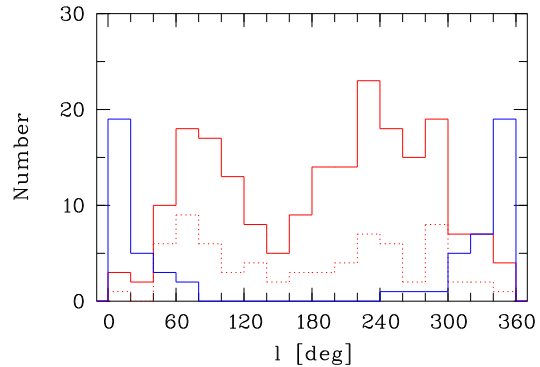}

\caption{\label{histl} Histogram showing the distribution of the clusters in our
sample along the Galactic Plane. In blue we show the globular clusters and in
red (solid line) all open clusters, while the dotted red line shows only
clusters with ages below one Gyr. The paucity of open clusters near the Galactic
Centre direction is a simple selection effect caused by the high star density
and thus low detection probability. The paucity between 120$^\circ < l
<$\,180$^\circ$ is not explained.} 

\end{figure}

Using the positions and determined distances to the clusters in our sample, we
can investigate their distribution in the Galactic Plane. In the left panel of
Fig.\,\ref{distrxy} we show the distribution of all clusters in the X-Y plane.
As blue triangles we plot the known globular clusters, while the open clusters
are plotted as red symbols (large circles for clusters with ages above 1\,Gyr,
small dots for clusters with ages below 1\,Gyr). The plot assumes a distance of
the Sun to the Galactic Centre of 8\,kpc. As expected one finds most of the
globular clusters concentrated towards the Galactic Centre. The open clusters are
mostly found near the Suns position. This is, however, a simple selection
effect, as the sample contains naturally only clusters near enough to be visible
in 2MASS data.

Two details of the spatial distribution of the open clusters are, however, worth
discussing in a bit more detail: 

i) A histogram of the distribution of Galactic Longitude values of the open
clusters in our sample (see Fig.\,\ref{histl}) reveals two regions with a
smaller than average number of clusters. This is a) the region near the Galactic
Centre ($\pm$\,60$^\circ$ away) and b) the Galactic Longitude range 120$^\circ <
l <$\,180$^\circ$. In the case of a) there are two reasons for this. Firstly,
the high star density towards the Galactic Centre prevents the detection and
identification of stellar clusters in this area (a selection effect when
establishing the sample) and secondly the fact that there are indeed fewer old
stellar clusters closer than the Sun to the Galactic Centre (see below). In the
case of b), there seems to be no obvious reason for the paucity of old open
clusters in this region compared to the same longitude range on the opposite
side of the Galactic Anticentre (i.e. the region 180$^\circ < l
<$\,240$^\circ$). One explanation could be one or several large, high extinction
molecular clouds in this direction, preventing the detection of clusters.
However, in the all sky extinction maps from Rowles \& Froebrich
\cite{2009MNRAS.395.1640R} there is no indication of such clouds. Furthermore,
the clusters which are detected in this longitude range, cover all distances
between 1 and 7\,kpc homogeneously, as well as extinction values between 0.0 and
0.4\,mag $A_K$. Finally, as can be seen in Fig.\,\ref{histl}, the effect is more
pronounced for clusters with ages above 1\,Gyr. Hence, this region either
suffers from an unknown selection effect, or indeed there are fewer than normal
old open clusters present in this part of the Galaxy.

ii) While our sample clearly contains a large number of clusters within 4\,kpc
from the Sun, there is a clear paucity of objects with distances to the Galactic
Centre less than that of the Sun. This is a fact already noticed by Friel
\cite{1995ARA&A..33..381F} and replicated since then. In our sample only 3\,\%
of the open clusters with ages above 1\,Gyr are closer than 5\,kpc to the
Galactic Centre (only 10\,\% are closer than 7\,kpc). In total 80\,\% of the old
open clusters are further away than the Sun from the Galactic Centre. This
clearly indicates that survival times of open clusters at smaller Galactocentric
distances than the one of the Sun are significantly shortened due to the
stronger tidal forces and the more frequent encounters with giant molecular
clouds (e.g. van den Bergh \& McClure \cite{1980A&A....88..360V}, Gieles et al.
\cite{2006MNRAS.371..793G, 2008IAUS..246..171G}).

We also analysed the distribution of clusters below and above the Galactic Plane
by fitting a Gaussian to the distribution. While the distribution of Galactic
Latitudes is slightly off-centre with the peak of the distribution at $b =
-0.6^\circ$ and a width of 7$^\circ$, the distribution of distances $Z$ to the
Galactic Plane is almost centred ($Z = -33$\,pc). In the right panel of
Fig.\,\ref{distrxy} we show the distribution of all clusters perpendicular to
the Galactic Plane. For clarity we zoom into the region where the open clusters
are situated. The older clusters seem to be much more widely distributed
perpendicular to the Galactic Plane than the younger objects. We hence analyse
the full width half maximum of the distributions for clusters older and younger
than 1\,Gyr. We find that the 137 clusters with ages equal to or above 1\,Gyr
have a scale height of 375\,pc. This agrees with earlier findings from e.g.
Janes \& Phelps \cite{1994AJ....108.1773J} utilising a much smaller sample of
clusters. In contrast, the clusters in our sample which are younger than 1\,Gyr
have a scale height of only 115\,pc. Even if our sample is inhomogeneous (see
below), this is a clear indication that the older clusters have either survived
longer due to their more inclined orbits, have been scattered into those, or
have been formed there as part of the thick disk. The different scale height are
not caused by selecting older clusters further away from the Galactic Centre
than the younger clusters. The average distance to the Galactic Centre is 8.9
and 9.4\,kpc for the younger and older clusters, respectively. Given the scatter
in these distances of about 2\,kpc, this difference is not enough to explain the
different scale heights.

\subsection{Reddening}

The extinction towards the clusters ranges from zero to more than 0.8\,mag
$A_K$, in a handful of cases. Generally there is a trend of the extinction
values with Galactic Longitude. Near to the Galactic Anticentre generally values
of 0.3\,mag $A_K$ are not exceeded and away from the Anticentre region we detect
more distant and reddened clusters. To account for the different distances we
investigated the distribution of the reddening per kiloparsec values of our
clusters. There are a handful of objects with more than 0.2\,mag $A_K$ per
kiloparsec distance, which are most likely clusters behind nearby giant
molecular clouds. The remaining clusters more or less show a homogeneous
distribution between zero and 0.1\,mag $A_K$/kpc. If we exclude clusters below
or above the scale height (outside the main disk) of the entire sample, we find
an average extinction for the entire sample of old open clusters of
0.70\,mag/kpc of optical extinction (conversion of $A_K$ into $A_V$ following
Mathis \cite{1990ARA&A..28...37M}).

\begin{figure}
\includegraphics[width=6.1cm,angle=-90]{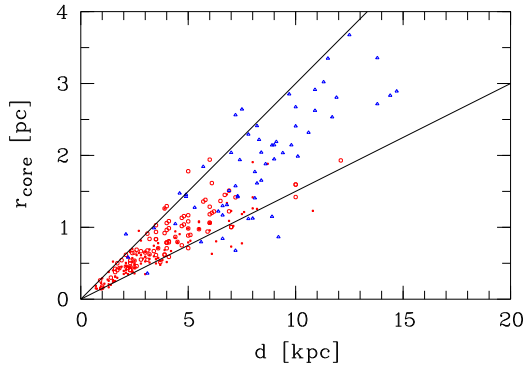}

\caption{\label{corr_d_r} The apparent correlation of distance and core radius
of our cluster sample, identified as one major selection effect in the FSR list.
Blue triangles are the globular clusters, while large and small red circles are
old (above one Gyr) and younger (less than one Gyr) open clusters. The two black
lines indicate the range for the correlation as given in Eq.\,\ref{eq_corr}.} 

\end{figure}

\subsection{Selection effects and Cluster Radii}

One so far not investigated issue of the FSR cluster candidate list is selection
effects. The cluster selection is of course influenced by the local background
star density and 2MASS completeness limit, as well as the distance to the
cluster and the extinction along the line of sight.

However, we seem to have found one further important selection bias in the FSR
sample. We investigate the core radius (in pc) of all clusters, determined from
the fit of the radial star density profile and the estimated distance from the
isochrone fitting (see Table\,\ref{properties} for the values of the individual
clusters). We find that there is a clear correlation of the cluster core radius
and its distance from the Sun (see Fig.\,\ref{corr_d_r}). In particular, if we
exclude the known globular clusters, the cluster core radii seem to follow the
relation

\begin{equation}\label{eq_corr}
r_{core}[pc] = ( 0.225 \pm 0.075 ) \cdot d[kpc]
\end{equation}
Hence, the FSR sample does neither contain compact distant clusters nor more
extended nearby objects. Both these effects are understandable when one looks
back at the cluster candidate selection procedure (see
Sect.\,\ref{dataanalysis}). Stars in compact distant clusters are simply not
resolved in the 2MASS data and hence might not have been picked up as local star
density enhancements, and/or the number of detectable cluster members is too
small to be identified as an old evolutionary sequence in the CMDs and CCDs.
Similarly, nearby clusters are more extended in the sky and are hence also not
picked up. Note that the average apparent cluster core radius of the old open
clusters in our sample is about 40", with a scatter of just $\pm$10". 

The identification of this selection effect in our cluster sample, also does not
allow us to study the evolution of cluster radius with age and/or position in
the Galaxy, as any trend might simply be caused by selection bias. If we try to
account for the selection effect, we can still obtain some tentative trends for
the cluster radii. i) more extended clusters seem to be found generally more
often at larger Galactocentric distance; ii) larger clusters seem to be found
generally at larger distances $Z$ from the Galactic Plane. This trend has
already been found by other authors (e.g. Janes et al.
\cite{1988AJ.....95..771J}), Tadross et al. \cite{2002NewA....7..553T},
Schilbach et al. \cite{	2006A&A...456..523S}).

\section{Conclusions}
\label{conclusions}

We have analysed the entire list of cluster candidates from Froebrich et al.
\cite{2007MNRAS.374..399F} by means of 2MASS photometry. We calculate more
accurate cluster positions and radii. For stars within each cluster we calculate
the membership probability based on a modified version of the
colour-colour-magnitude approach by Bonatto \& Bica \cite{2007MNRAS.377.1301B}.
A by eye inspection of 2MASS colour-magnitude and colour-colour diagrams of high
probability members has been used to identify 269 candidates for old clusters
in the FSR sample.

Our sample of clusters contains 63 known globular clusters, 147 known open
clusters with literature parameters, 27 known open clusters without known
parameters and 32 previously unknown objects. We use isochrones from Girardi et
al. \cite{2002A&A...391..195G} to determine the age, distance and reddening for
each cluster homogeneously from the 2MASS photometry. Our sample has a mean age
of 1\,Gyr and 80\,\% of the clusters are older than 500\,Myrs.

The distribution of open clusters in the Galaxy shows that 80\,\% of the
clusters with ages above 1\,Gyr are further away than 7\,kpc from the Galactic
Centre, strengthening earlier findings e.g. from Friel
\cite{1995ARA&A..33..381F}. Furthermore, the scale height of these old clusters
is with 375\,pc more than three times as large as the scale height of the
younger (less than 1\,Gyr) fraction in our sample, which we found to be 115\,pc.
This is still about twice as large as the scale height of young star clusters
which is about 55\,pc (Friel \cite{1995ARA&A..33..381F}). 

The large sample of clusters also allows us to investigate the general
interstellar extinction for objects mostly not associated with Giant Molecular
Clouds. For our sample of old clusters we find an average interstellar optical
extinction of 0.70\,mag/kpc.

We also identify a main selection effect in the FSR cluster sample, besides
distance and reddening. An investigation of the cluster radii shows that the
sample contains mostly clusters which have a distance to radius ratio in the
range 3.3 to 6.7\,$\cdot$\,10$^3$. This is caused by the fact that the sample
seems to be biased towards clusters with an apparent projected core radius of
40"\,$\pm$\,10". 

Finally we find that there seems to be a significantly smaller than average
number of old open clusters in the longitude range 120$^\circ < l
<$\,180$^\circ$, which cannot be explained by any of the sample selection
effects. The reason for this paucity is unknown.

Large improvements can be expected in the near future for this kind of work, as
the UKIDSS GPS and Vista VVV surveys can be utilised. Their increase in limiting
magnitude and in particular spatial resolution compared to 2MASS will allow us
to improve on the parameter determinations, as well as to build up a cluster
sample with less selection effects, e.g. circumventing the core radius selection
effect.

\section*{acknowledgments}

S.Schmeja acknowledges funding by the Deutsche Forschungsgemeinschaft (DFG)
through grant SCHM 2490/1-1. This publication makes use of data products from
the Two Micron All Sky Survey, which is a joint project of the University of
Massachusetts and the Infrared Processing and Analysis Center/California
Institute of Technology, funded by the National Aeronautics and Space
Administration and the National Science Foundation. This research has made use
of the SIMBAD database, operated at CDS, Strasbourg, France. This research has
made use of the WEBDA database, operated at the Institute for Astronomy of the
University of Vienna.

\clearpage\newpage

\begin{appendix}

\onecolumn

\begin{landscape}

\section{Main Table}

\begin{longtable}{l|rr|rr|rrrrr|rrrr|rc|p{2cm}}

\caption{\label{properties} Summary table of the properties of the old stellar
clusters analysed in this paper. We list the FSR catalogue ID, the coordinates,
BIC value and core radius, the cluster properties as used in the isochrone fit in
this work, the WEBDA or literature cluster parameter, the separation and
classification of the known clusters in SIMBAD, and finally other common names
for the clusters. The parameters for the isochrone fit used in this paper include
the distance to the cluster, its age, the K-band extinction, the metallicity and
the reddening law. The WEBDA and literature values are the distance of the
cluster, its age, the reddening and metallicity. To plot the literature
isochrone for the known globular clusters we use an age of 12\,Gyrs. For
clusters marked with a $^*$, please read the notes in Appendix\,\ref{notes} for
details on the isochrone fits.} \\

\noalign{\smallskip}\hline\noalign{\smallskip} 
FSR & \multicolumn{2}{|c|}{Coordinates} & & & \multicolumn{5}{|c|}{this paper} & \multicolumn{4}{|c|}{Literature$^1$} & \multicolumn{2}{|c|}{SIMBAD} & other \\
ID & l & b & BIC & $r_{core}$ & d & log(age) & A$_{\rm K}$ & [M/H] & C$_{\rm HK}$ & d & log(age) & E(B-V) & [M/H] & r & Class & Names \\
  & [deg] & [deg] & & [pc] & [kpc] & [yr] & [mag] & [dex] &  & [kpc] & [yr] & [mag] & [dex] & ["] &  &  \\
\noalign{\smallskip}\hline\noalign{\smallskip} 
\endfirsthead
\noalign{\smallskip}\hline\noalign{\smallskip} 
\multicolumn{17}{l}{{\bfseries \tablename\ \thetable{} -- continued from previous page}} \\
\noalign{\smallskip}\hline\noalign{\smallskip} 
FSR & \multicolumn{2}{|c|}{Coordinates} & & & \multicolumn{5}{|c|}{this paper} & \multicolumn{4}{|c|}{Literature$^1$} & \multicolumn{2}{|c|}{SIMBAD} & other \\
ID & l & b & BIC & $r_{core}$ & d & log(age) & A$_{\rm K}$ & [M/H] & C$_{\rm HK}$ & d & log(age) & E(B-V) & [M/H] & r & Class & Names \\
  & [deg] & [deg] & & [pc] & [kpc] & [yr] & [mag] & [dex] &  & [kpc] & [yr] & [mag] & [dex] & ["] &  &  \\
\noalign{\smallskip}\hline\noalign{\smallskip} 
\endhead
\noalign{\smallskip}\hline\noalign{\smallskip} 
\multicolumn{17}{r}{Continued on next page} \\ 
\noalign{\smallskip}\hline\noalign{\smallskip} 
\endfoot
\noalign{\smallskip}\hline\noalign{\smallskip}
\endlastfoot
0003          &   0.069 & $-$17.2926 & $-$289.4 & 2.4 &  8.20 & 10.00 & 0.05 & $-$1.28 & 1.97 &  8.2 &   --- & 0.04 & $-$1.12 &  23 & GlC     & NGC\,6723                                     \\                           
0004          &   0.130 & $+$11.0276 & $-$146.6 & 1.6 &  8.20 & 10.20 & 0.23 & $-$2.28 & 1.63 &  8.2 &   --- & 0.59 & $-$2.05 &  17 & GlC     & NGC\,6287                                     \\                           
0006          &   0.982 &  $+$8.0110 &   $-$2.5 & 2.1 &  9.00 & 10.20 & 0.30 & $-$1.28 & 1.74 &  9.0 &   --- & 0.89 & $-$1.17 &  43 & GlC     & NGC\,6325                                     \\                           
0007          &   1.541 & $-$11.3652 &  $-$40.3 & 1.9 &  9.00 & 10.20 & 0.03 & $-$0.68 & 1.67 &  9.0 &   --- & 0.09 & $-$0.96 &  39 & GlC     & NGC\,6652                                     \\                           
0008          &   1.729 & $-$10.2619 &  $-$36.5 & 2.3 &  7.80 & 10.20 & 0.06 & $-$0.68 & 1.67 &  7.8 &   --- & 0.17 & $-$0.71 &  35 & GlC     & M\,69, NGC\,6637                                 \\                        
\end{longtable}

Table notes:

$^1$ Data for the globular clusters has been taken from Harris
\cite{1996AJ....112.1487H} and the data for the open clusters from the WEBDA
database, except when indicated otherwise by a footnote.

\end{landscape}
\clearpage\newpage
\twocolumn 

\section{Notes on individual clusters}\label{notes}

In the following we will discuss details to some clusters. These are either the
objects with parameters determined here for the first time, or clusters with
literature values that do not fit the 2MASS data. If the clusters is known
already, then its other common names are given beside the FSR number.

{\bf FSR\,0039:} This seems to be a newly discovered old open cluster just
4.6\,kpc from the Galactic Centre. The sequence in the CMD and CCD suggest we
see a number of red giants, including red clump stars. There is no detectable
main sequence in the 2MASS data, hence the age of 1\,Gyr is an upper limit. The
cluster seems highly reddened with $A_K = 1.22$\,mag, but at a distance of
3.5\,kpc and a Galactic Longitude of just 10$^\circ$ this is no surprise.

{\bf FSR\,0050:} The CMD and CCD sequence suggest that this is an about 5\,Gyr
old open cluster. There are not many red giant stars to determine the properties
accurately, however the isochrone fit seems to fit the top of the main sequence
the giants well. If confirmed, this object as well would be an inner Galaxy old
open cluster with only 5.6\,kpc distance to the Galactic Centre.

\clearpage\newpage

\section{Colour-Magnitude and Colour-Colour Diagrams}
\label{CMD}
\clearpage\newpage

\begin{figure}\includegraphics[width=6.1cm,angle=-90]{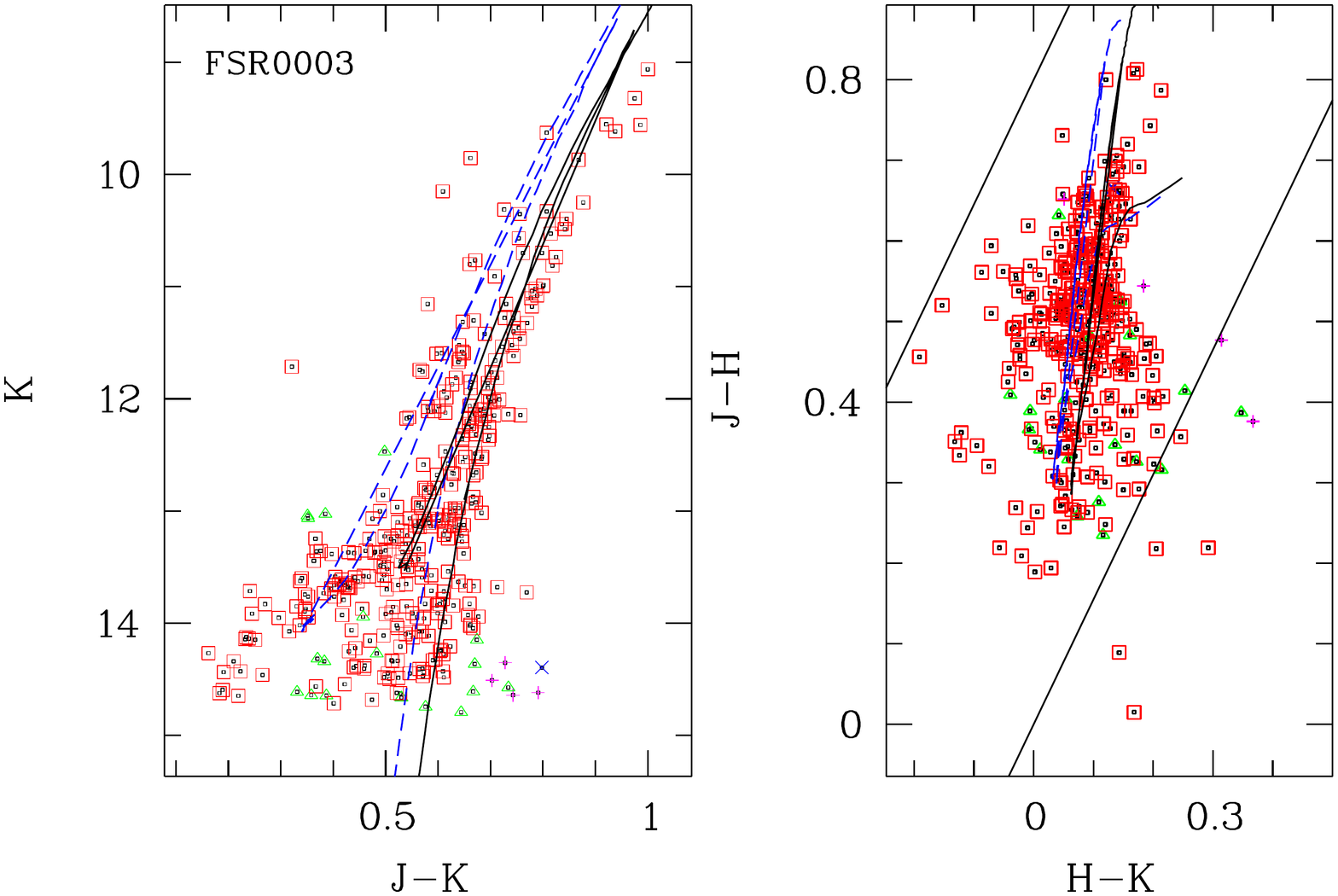}\caption{\label{cmd0003} Example of our colour-magnitude
(left) and colour-colour diagrams for the cluster FSR\,0003 (NGC\,6723). Red squares are stars with $P$\,$>$\,80\,\%, green triangles are
stars with 60\,\%\,$<$\,$P$\,$<$\,80\,\%, pink $+$-signs are stars with 40\,\%\,$<$\,$P$\,$<$\,60\,\%, blue crosses are stars with
20\,\%\,$<$\,$P$\,$<$\,40\,\% and black dots are stars with $P$\,$<$\,20\,\%. Overplotted in black is the best fitting isochrone (see
Table\,\ref{properties} for the parameters). The two vertical lines enclose the reddening band for stellar atmospheres.} \end{figure}

\end{appendix}

\label{lastpage}

\end{document}